\documentclass[12pt]{article}
\usepackage{graphicx}
\begin{document}

\begin{center}

{\LARGE \bf Quark-Parton Phase Transitions \\[1ex]
and the \\[1.5ex] Entropy Problem in Quantum Mechanics}

\vspace{15mm}

Y. S. Kim\footnote{electronic mail: yskim@umd.edu} \\
Department of Physics, University of Maryland, \\
College Park, Maryland 20742, U.S.A.

\end{center}

\vspace{10mm}
\begin{abstract}

Since Feynman proposed his parton model in 1969, one of the most
pressing problems in high-energy physics has been whether
partons are quarks.  It is shown that the quark model and the
parton model are two different manifestations of one covariant
entity.  The nature of transition from the confined quarks to
plasma-like partons is studied in terms of the entropy and
temperature coming from the time-separation variable.  According
to Einstein, the time-separation variable exists wherever there is
a spatial separation, but it is not observed in the present form of
quantum mechanics.  The failure to observe this variable causes an
increase in entropy.
\end{abstract}

\vspace{60mm}

 \noindent Presented at the Akhiezer Memorial
 Conference on Quantum Electrodynamics and Statistical Physics
 (Kharkov, Ukraine, 2001),
 and published in the Problems of Atomic Science and Technology,
 Special Issue dedicated to the 90th Birthday Anniversary of A. I.
 Akhiezer, No. 6 (1), 149--153 (2001).

\vspace{3mm}

\noindent http://arxiv.org/abs/quant-ph/0201010

\newpage

\section{Introduction}\label{intro}

In 1969, Feynman proposed his parton model for hadrons moving with
speed close to that of light~\cite{fey69}.  He observed that the
hadron appears as a collection of infinite number of partons.  Since
the partons appear to have properties quite different from those of
the quarks, one of the most pressing puzzles in high-energy physics
has been whether the partons are quarks, or whether the quark model
and the parton model are two different manifestations of one covariant
formalism.

In 1970, at the April meeting of the American Physical Society held
in Washington, DC (U.S.A.), Feynman gave an invited talk on a model
of hadrons.  His talk was published in a paper by Feynman, Kislinger
and Ravndal in 1971~\cite{fkr71}.  There, the authors attempted to
construct a covariant model for hadrons consisting of quarks joined
together by an oscillator force.  They indeed formulated a
Lorentz-invariant oscillator equation.  They also worked out the
degeneracies of the oscillator states which are consistent with
observed mesonic and baryonic mass spectra.  However, their wave
functions are not normalizable in the space-time coordinate system.
They never considered the question of covariance.

In his 1972 book on statistical mechanics~\cite{fey72}, Feynman says
{\it When we solve a quantum-mechanical problem, what we really do
is divide the universe into two parts - the system in which we
are interested and the rest of the universe.  We then usually act as
if the system in which we are interested comprised the entire universe.
To motivate the use of density matrices, let us see what happens when
we include the part of the universe outside the system}.  Feynman's
rest of the universe has been studied in detail in terms of two
coupled oscillators~\cite{hkn99ajp}.

In this report, we combine these three components of Feynman's research
efforts to show that the quark and parton models are indeed two
different manifestations of the same covariant entity.  In order to
achieve this purpose, we fix up first the mathematical deficiencies of
the paper of Feynman {\it et al.}~\cite{fkr71}.
The idea is to construct a harmonic oscillator wave function which can
be Lorentz-boosted.  We can first see whether the wave function is
applicable to the quark model when the hadron is slow, and then see
whether the same wave function describes the parton model when the
hadron is boosted to an infinite-momentum frame.  The 1971 paper by
Feynman {\it et al.}~\cite{fkr71} contains very serious mathematical
flaws, but they have been all cleaned up within the framework of
Wigner's little groups which dictate the internal space-time
symmetries relativistic particles~\cite{wig39,knp86}.

This covariant formulation solves the covariance problem.  However,
since we live in the three-dimensional world, it is possible that
we miss something in the four-dimensional world.  The time-separation
variable between the quarks is a case in point. In non-relativistic
quantum mechanics, the Bohr radius is spacial separation between the
quarks (or proton and electron).  According to Einstein, there must
be a time separation between the quarks, since otherwise the world
will not be covariant.

Since we are not dealing with this time-separation variable in the
present form of quantum mechanics, the failure to measure it leads
to an increase in entropy~\cite{fey72}.  In this report, we show
that this entropy allows us to define the phase transition between
the confined phase of the quark model and the plasma phase of the
parton model.

In Sec.~\ref{covham}, we introduce the covariant harmonic oscillator
formalism with normalizable wave functions which can be Lorentz
boosted.  In Sec.~\ref{parton}, we use the oscillator wave function
to solve the quark-parton puzzle.  In Sec.~\ref{entro}, we deal with
the problems arising from measuring of four-dimensional physics in
the three-dimensional world.  The entropy plays a major role.

\section{Covariant Harmonic Oscillators}\label{covham}

\begin{figure}
\centerline{\includegraphics[scale=0.7]{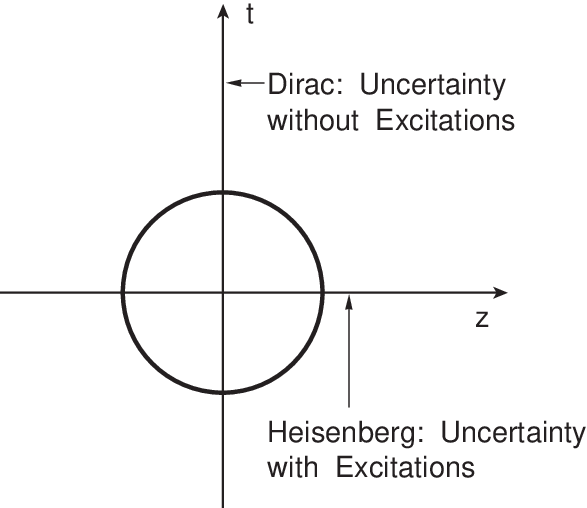}}
\vspace{5mm}
\caption{Present form of quantum mechanics.  There are excitations
along the space-like dimensions, but there are no excitations along
the time-like direction.  However, there still is a time-energy
uncertainty relation.  We call this Dirac's c-number time-energy
uncertainty relation.  It is very important to note that this
space-time asymmetry is quite consistent with the concept of
covariance.}\label{f.quantum}
\end{figure}

Let us consider a hadron consisting of two quarks.
Then there is a Bohr-like radius measuring the space-like separation
between the quarks.  There is also a time-like separation between the
quarks, and this variable becomes mixed with the longitudinal spatial
separation as the hadron moves with a relativistic speed.  While
there are no quantum excitations along the time-like direction,
there is the time-energy uncertainty relation which allows
quantum transitions.  It is possible to accommodate these aspects within
the framework of the present form of quantum mechanics.  The uncertainty
relation between the time and energy variables is the c-number
relation~\cite{dir27}, which does not allow excitations along the
time-like coordinate, as illustrated in Fig.~\ref{f.quantum}

\begin{figure}[thb] 
\centerline{\includegraphics[scale=0.6]{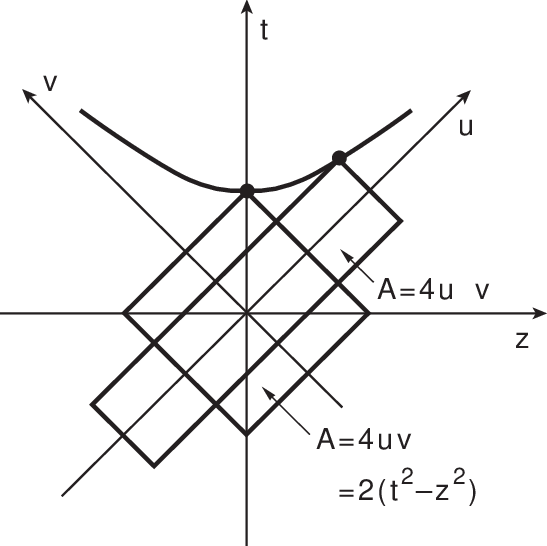}}
\vspace{5mm}
\caption{Lorentz boost in the light-cone coordinate system.  The boost
expands one of the light-cone axes while contracting the other.
figure.}\label{f.licone}
\end{figure}

For a hadron consisting of two quarks, we can consider their space-time
positions $x_{a}$ and $x_{b}$, and use the variables
\begin{equation}
X = (x_{a} + x_{b})/2 , \qquad x = (x_{a} - x_{b})/2\sqrt{2} .
\end{equation}
The four-vector $X$ specifies where the hadron is located in space and
time, while the variable $x$ measures the space-time separation between
the quarks.

Since the three-dimensional oscillator differential equation is
separable in both spherical and Cartesian coordinate systems,
the wave function consists of Hermite polynomials of $x, y$, and $z$.
If the Lorentz boost is made along the $z$ direction, the $x$ and $y$
coordinates are not affected, and can be temporarily dropped from the
wave function.  Along the space-like longitudinal direction,
there are excitations.  On the other hand, along the time-like direction,
there is an uncertainty relation even though there are no excitations.
The covariant harmonic oscillator formalism accommodates this space-time
asymmetry~\cite{knp86}.

\begin{figure}[thb] 
\centerline{\includegraphics[scale=0.7]{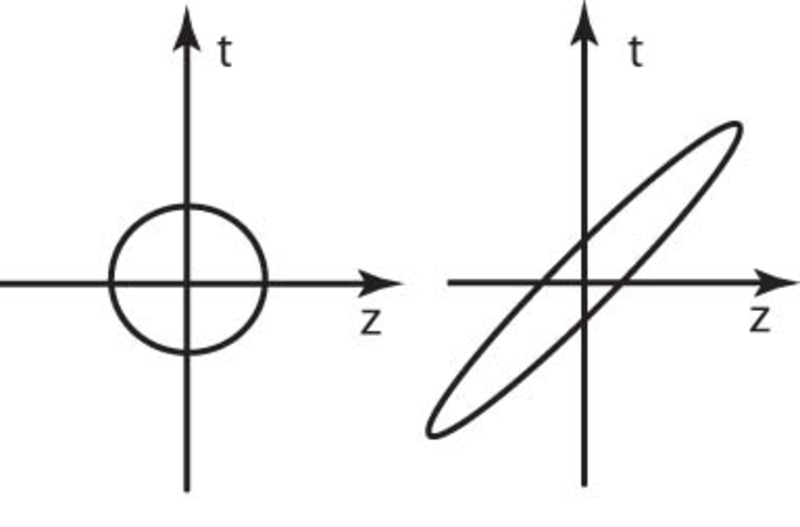}}
\vspace{2mm}
\caption{Effect of the Lorentz boost on the space-time wave function.
The circular space-time distribution at the rest frame becomes
Lorentz-squeezed to become an elliptic distribution.}\label{f.elli}
\end{figure}


However, since we are interested here only in
Lorentz-boost properties of the wave function, we restrict ourselves
to the ground-state wave function.  The wave function then can be
written as
\begin{equation} \label{12}
\psi(z,t) = \left({1\over \pi }\right)^{1/2}
\exp\left\{-{1\over 2}\left(z^{2} + t^{2} \right) \right\} ,
\end{equation}
which accommodates the uncertainty relations along the longitudinal
and time-like directions.

The expression given in Eq.(\ref{12}) is not Lorentz-invariant.  It is
covariant. This wave function describes the present form of quantum
mechanics if the time-separation variable is factored out,
integrated out, or ignored.
However, the time-separation variable is absolutely needed when we
consider Lorentz covariance.  The question is whether the above wave
function can describe the parton model when it is boosted to an
infinite-momentum limit.

It is convenient to use the light-cone variables to describe Lorentz
boosts.  The light-cone coordinate variables are
\begin{equation}
u = (z + t)/\sqrt{2} , \qquad v = (z - t)/\sqrt{2} .
\end{equation}
In terms of these variables, the Lorentz boost along the $z$
direction,
\begin{equation}
\pmatrix{z' \cr t'} = \pmatrix{\cosh \eta & \sinh \eta \cr
\sinh \eta & \cosh \eta } \pmatrix{z \cr t} ,
\end{equation}
takes the simple form
\begin{equation}\label{lorensq}
u' = e^{\eta } u , \qquad v' = e^{-\eta } v ,
\end{equation}
where $\eta$ is the boost parameter and is $\tanh^{-1}(v/c)$.
Indeed, the $u$ variable becomes expanded while the $v$ variable becomes
contracted.  This is the squeeze mechanism illustrated discussed
extensively in the literature~\cite{kn73,knp91}.  This squeeze
transformation is also illustrated in Fig.~\ref{f.licone}.

Thus, one way to combine quantum mechanics with relativity is to
incorporate Fig.~\ref{f.quantum} into Fig.~\ref{f.licone}, and produce
the elliptic deformation illustrated in Fig.~\ref{f.elli}.
If the system is boosted, the wave function becomes
\begin{equation}\label{15}
\psi_{\eta }(z,t) = \left({1 \over \pi }\right)^{1/2}
\exp \left\{-{1\over 2}\left(e^{-2\eta }u^{2} +
e^{2\eta }v^{2}\right)\right\} .
\end{equation}
We note here that the transition from Eq.(\ref{12}) to Eq.(\ref{15})
is a squeeze transformation.  The wave function of Eq.(\ref{12}) is
distributed within a circular region in the $u v$ plane, and thus in
the $z t$ plane.  On the other hand, the wave function of Eq.(\ref{15})
is distributed in an elliptic region.  This ellipse is a ``squeezed''
circle with the same area as the circle, as is illustrated in
Fig.~\ref{f.elli}.

\section{Feynman's Parton Picture}\label{parton}

In 1969~\cite{fey69} Feynman made the following systematic observations
on hadrons moving with speed close to that of light.

\begin{itemize}

\item[ a).] The picture is valid only for hadrons moving with velocity
     close to that of light.

\item[ b).] The interaction time between the quarks becomes dilated,
  and partons behave as free independent particles.

\item[ c).] The momentum distribution of partons becomes widespread as
  the hadron moves very fast.

\item[ d).] The number of partons seems to be infinite and much larger
  than that of quarks.

\end{itemize}

\noindent  These observations constitute Feynman's parton picture.
Because the hadron is believed to be a bound state of two or
three quarks, each of the above phenomena appears as a paradox,
particularly b) and c) together.

If the quarks have the four-momenta $p_{a}$ and $p_{b}$, we can
construct two independent four-momentum variables~\cite{fkr71}
\begin{equation}
P = p_{a} + p_{b} , \qquad q = \sqrt{2}(p_{a} - p_{b}) .
\end{equation}
The four-momentum $P$ is the total four-momentum and is thus the hadronic
four-momentum.  $q$ measures the four-momentum separation between the quarks.

\begin{figure}
\centerline{\includegraphics[scale=0.43]{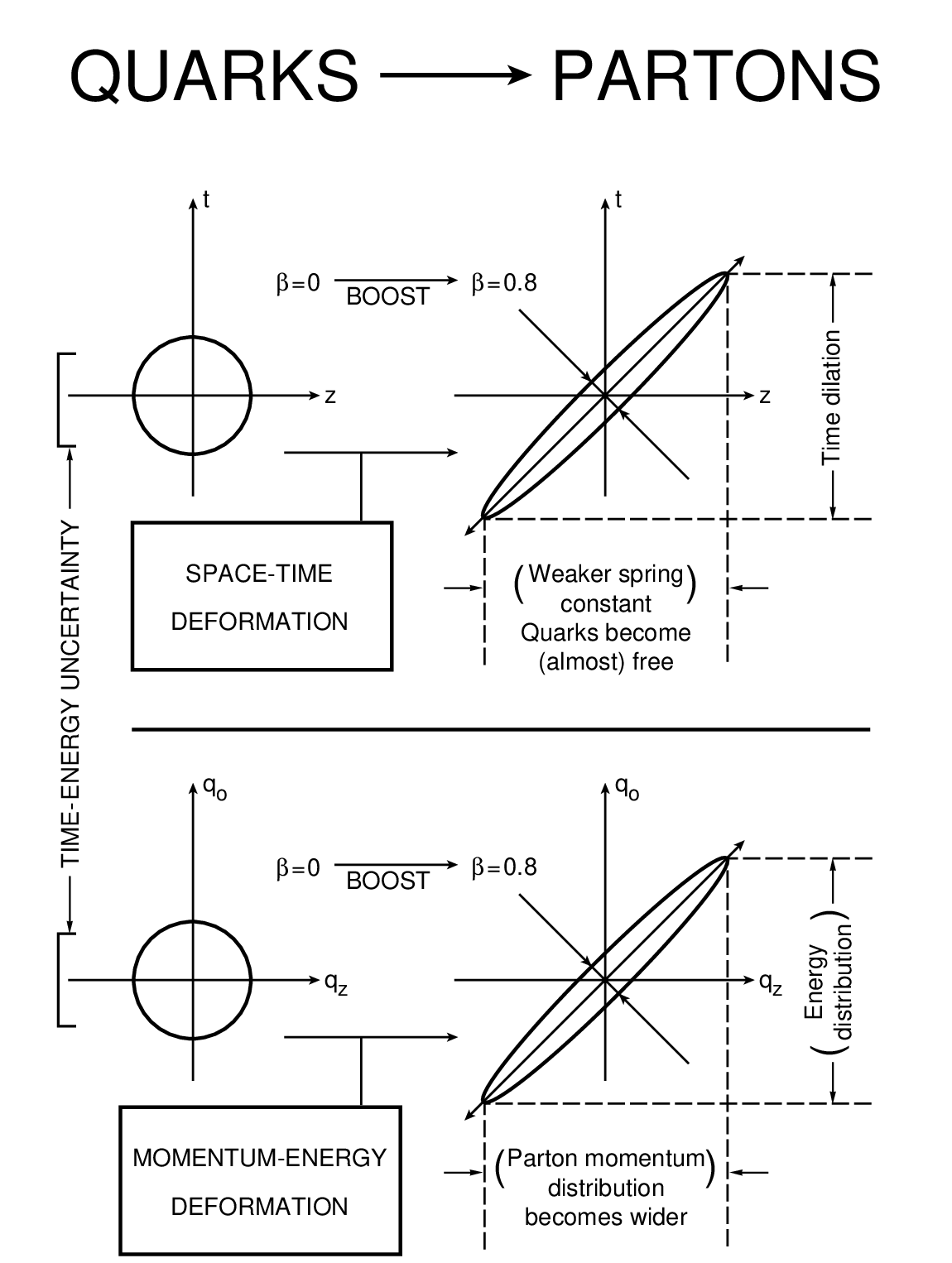}}
\vspace{5mm}
\caption{Lorentz-squeezed space-time and momentum-energy wave functions.
As the hadron's speed approaches that of light, both wave functions
become concentrated along their respective positive light-cone axes.
These light-cone concentrations lead to Feynman's parton
picture.}\label{f.parton}
\end{figure}
Since we are using here the harmonic oscillator, the mathematical form
`of the above momentum-energy wave function is identical to that of the
space-time wave function, and its transformation properties are the same.
The Lorentz squeeze properties of these wave
functions are also the same, as are indicated in Fig.~\ref{f.parton}.
When the hadron is at rest with $\eta = 0$, both wave functions behave
like those for the static bound state of quarks.  As $\eta$ increases,
the wave functions become continuously squeezed until they become
concentrated along their respective positive light-cone axes.  Indeed,
this figure provides the answer to the quark-parton puzzle~\cite{knp86}.

\begin{figure}[thb] 
\centerline{\includegraphics[scale=0.5]{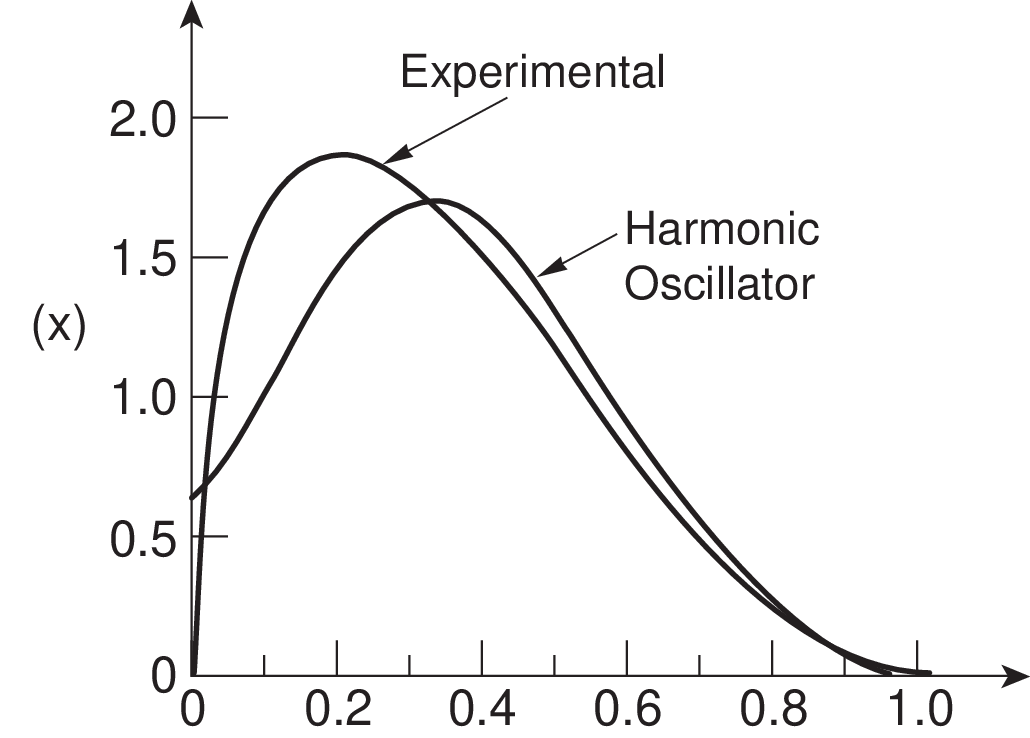}}
\caption{Parton distribution.  It is possible to calculate the parton
distribution from the Lorentz-boosted oscillator wave function.
This theoretical curve is compared with the experimental
curve.}\label{f.hussar}
\end{figure}

The question then is whether the elliptic deformations given in
Fig.~\ref{f.parton} produce any quantitative results which can be
compared with what we measure in laboratories.
Indeed, according to Hussar's calculation~\cite{hussar81},
the Lorentz-boosted oscillator wave function produces a reasonably
accurate parton distribution, as indicated in Fig.~\ref{f.hussar}

\section{Entropy Problems}\label{entro}

The covariant harmonic oscillator formalism presented in Sec.~\ref{covham}
produces the Lorentz squeeze property summarized in Fig.~\ref{f.parton}.
This figure tells us that the quark model and the parton model are two
different manifestations of one covariant formulation.  In this figure,
the time-separation variable plays the essential role.  However, we
are not able to deal with this variable in the present form of quantum
mechanics.

If there is a physical which we cannot measure, the variable certainly
belongs to Feynman's rest of the universe~\cite{fey72,hkn99ajp}.  Then
there is a well-defined procedure to deal with this problem: construct
a density matrix from the wave function and integrate over the variable
which we do not observe.  In the present case, the variable we do not
observe is the time-separation variable.  This
process leads to an increase in entropy~\cite{kiwi90pl}.  It is
straight-forward to calculate this entropy~\cite{hkn99ajp}, and the
result is
\begin{equation}\label{entro2}
S = 2 \left\{(\cosh^{2}\eta) \ln(\cosh\eta) -
   (\sinh^{2}\eta)\ln(\sinh\eta)\right\} .
\end{equation}
This form is identical to the entropy caused by thermally
excited harmonic oscillators, if we write
\begin{equation}\label{temp}
\tanh^{2}(\eta) = \exp{\left({-\hbar\omega \over kT }\right)} .
\end{equation}
The entropy of Eq.(\ref{entro2}) takes the
form~\cite{kiwi90pl,hkn89pl}
\begin{equation}
S ={\hbar\omega/kT \over \exp{(\hbar\omega/kT)} - 1}
- \ln\left[1 - \exp{\left({-\hbar\omega \over kT}\right)} \right] .
\end{equation}

\begin{figure}[thb] 
\centerline{\includegraphics[scale=0.9]{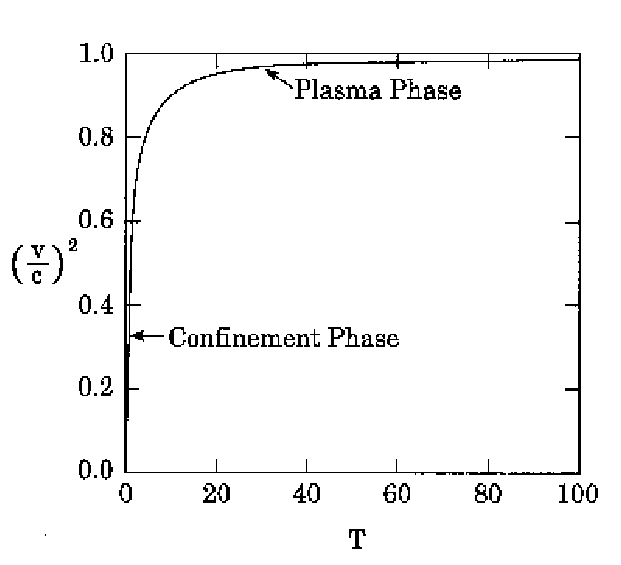}}
\caption{The hadronic velocity versus the hadronic temperature given
in Eq.(\ref{temp}).  Here we used the unit system where
$\hbar\omega/k = 1$, and $\tanh\eta = v/c$.}\label{f.qplas}
\end{figure}

Let us go back to Eq.(\ref{temp}).  The $(velocity)^{2}$ is plotted
against the temperature in Fig.~\ref{f.qplas}.  Its behavior makes
a sudden change as the temperature rises.  If the hadronic velocity
is low, the temperature is relatively insensitive to the velocity,
but for high velocities, it is in the other way around.  We can use
this behavior to tell the difference between the confinement phase
of the quarks and the plasma phase of the partons.

\begin{thebibliography}{99}

\bibitem{fey69}
R. P. Feynman, in {\it High Energy Collisions}, Proceedings of the
Third International Conference, Stony Brook, New York, edited by
C. N. Yang {\it et al}. (Gordon and Breach, New York, 1969).

\bibitem{fkr71}
R. P. Feynman, M. Kislinger, and F. Ravndal, Phys. Rev. D {\bf 3}, 2706
(1971).

\bibitem{fey72}
R. P. Feynman, {\it Statistical Mechanics} (Benjamin,
Reading, MA, 1972).

\bibitem{hkn99ajp}
D. Han, Y. S. Kim, and M. E. Noz, Am. J. Phys. {\bf 67}, 61 (1999).

\bibitem{wig39}
E. P. Wigner, Ann. Math. {\bf 40}, 149 (1939).

\bibitem{knp86}
Y. S. Kim and M. E. Noz, {\it Theory and Applications of the Poincar\'e
Group} (Reidel, Dordrecht, 1986).

\bibitem{dir27}
P. A. M. Dirac, Proc. Roy. Soc. (London) {\bf A114}, 243 and 710 (1927).

\bibitem{kn73}
Y. S. Kim and M. E. Noz, Phys. Rev. D {\bf 8}, 3521 (1973).

\bibitem{knp91}
Y. S. Kim and M. E. Noz, {\it Phase Space Picture of Quantum
Mechanics} (World Scientific, Singapore, 1991).

\bibitem{hussar81}
P. E. Hussar, Phys. Rev. D {\bf 23}, 2781 (1981).

\bibitem{kiwi90pl}
Y. S. Kim and E. P. Wigner, Phys. Lett. A {\bf 147}, 343 (1990).

\bibitem{hkn89pl}
D. Han, Y. S. Kim, and M. E. Noz, Phys. Lett. A {\bf 144}, 111 (1989).

\end{thebibliography}
\end{document}